\documentclass[review,12pt,numbers]{elsarticle}

\graphicspath{ {./figures/}{./charts/} } 

\fontfamily{ptm} \selectfont

\usepackage{setspace} 
\usepackage[left=2cm, right=2cm, top=2cm]{geometry} 

\usepackage{float}
\usepackage[section]{placeins}

\usepackage{amsmath, amsfonts, amsthm, amssymb} 
\usepackage{upgreek, bm, bbm} 

\usepackage[table]{xcolor} 
\usepackage{booktabs} 
\usepackage{multirow} 
\usepackage{rotating} 

\usepackage[caption=false]{subfig} 
\usepackage{caption}

\usepackage[colorlinks=true]{hyperref} 
\usepackage[capitalise]{cleveref}
\usepackage{soul}
\usepackage[normalem]{ulem}
\usepackage{lipsum} 

\usepackage[defaultcolor=red,authormarkup=none,draft]{changes} 
\definechangesauthor[name={Riccardo}, color=red]{RA1} 
\definechangesauthor[name={Riccardo}, color=magenta]{RA2} 
\definechangesauthor[name={Anyone}, color=orange]{MOV} 

\newcommand{\X}{\mathbf{X}}
\newcommand{\x}{\mathbf{x}}
\newcommand{\F}{\mathbf{F}}
\newcommand{\C}{\mathbf{C}}
\newcommand{\N}{\mathbf{N}}
\newcommand{\n}{\mathbf{n}}
\newcommand{\de}{\mathrm{d}}
\newcommand{\I}{\mathbf{I}}
\renewcommand{\b}{\mathbf{b}}
\newcommand{\stiffc}{ {\scriptstyle{\mathbb{C}}} }
\newcommand{\Nip}{\mathbf{N}_{\mathrm{ip}}}
\newcommand{\Mi}{\mathbf{M}_{i}}
\newcommand{\Mipi}{\mathbf{M}_{\mathrm{ip},i}}
\newcommand{\Mopi}{\mathbf{M}_{\mathrm{op},i}}
\newcommand{\ip}{\mathrm{ip}}
\newcommand{\op}{\mathrm{op}}
\newcommand{\eend}{\:}

\newcommand{\parr}[1]{\left(#1\right)} 
\newcommand{\pars}[1]{\left[#1\right]} 
\newcommand{\Ifour}[1]{I_{4,#1}} 
\newcommand{\E}[1]{\mathbf{E}_{#1}}
\newcommand{\norm}[1]{\left\lVert#1\right\rVert} 

\DeclareMathOperator\erf{erf} 
\DeclareMathOperator\symm{symm} 

\journal{CMAME}

\begin{document}

\begin{frontmatter}

	\title{{\bf{A Discrete Fiber Dispersion Model with Octahedral Symmetry Quadrature for Applications in Skin Mechanics}}}

	\author[a]{Riccardo Alberini}
	\author[b]{Michele Terzano}
	\author[b,c]{Gerhard A. Holzapfel}
	\author[a]{Andrea Spagnoli\corref{cor1}}
	\address[a]{Department of Engineering and Architecture, University of Parma, Parma, Italy}
	\address[b]{Institute of Biomechanics, Graz University of Technology, Graz, Austria}
	\address[c]{Department of Structural Engineering, Norwegian University of Science and Technology, Trondheim, Norway}
	\cortext[cor1]{Corresponding author\\
		Tel.: +39 0521 905927\\
		\textit{e-mail}: spagnoli@unipr.it}

	\begin{abstract}
		Advanced simulations of the mechanical behavior of soft tissues frequently rely on structure-based constitutive models, including smeared descriptions of collagen fibers. Among them, the so-called Discrete Fiber Dispersion (DFD) model is based on a discrete integration of the fiber-strain energy over all the fiber directions. In this paper, we recall the theoretical framework of the DFD model, including a derivation of the stress and stiffness tensors required for the finite element implementation. Specifically, their expressions for incompressible plane stress problems are obtained. The use of a Lebedev quadrature, built exploiting the octahedral symmetry, is then proposed, illustrating the particular choice adopted for the orientation of the integration points. Next, the convergence of this quadrature scheme is assessed by means of three numerical benchmark tests, highlighting the advantages with respect to other angular integration methods available in the literature. Finally, we propose as applicative example a simulation of Z-plasty, a technique commonly used in reconstructive skin surgery, considering multiple geometrical configurations and orientations of the fibers. Results are provided in terms of key mechanical quantities relevant for the surgical practice. 
	\end{abstract}


\end{frontmatter}


\section{Introduction}
In recent years, the development of structure-based mechanical models has gained significant attention in computational mechanics, particularly for materials with a complex internal microstructure such as soft collagenous tissues. These models aim to capture the non-linear and anisotropic material behavior more accurately with respect to purely phenomenological formulations, by accounting for the non-uniform dispersion of the fibers within the material~\cite{Benitez2017}. Tissues such as arterial walls, heart valves, gastric muscle, and skin are mainly constituted by a dispersion of wavy collagen fibers embedded in a ground isotropic substance~\cite{Ayyalasomayajula2019, Sadeghinia2022, Holzer2024, Alberini2024a}. The orientation distribution of the fibers in the three-dimensional space is given by a Probability Density Function (PDF), which can be included in the constitutive model according to two main approaches: the Generalized Structure Tensor (GST) and the Angular Integration (AI) approaches~\cite{Holzapfel2019}. In the GST approach, the PDF is used to calculate a structure tensor representative of the fiber distribution, which is then used in the strain-energy function to determine a pseudo-invariant accounting for the anisotropy. Based on a specific shape of the PDF, the structure tensor can be computed in a closed form, without requiring further integration during the analysis. For this reason, models based on this approach, such as those proposed by \citet{Merodio2005, Gasser2006, Holzapfel2015} to name a few, became popular due to their computational efficiency~\cite{Benitez2017}.

A downside of these models is that they cannot exclude the contribution of compressed fibers along individual directions without compromising the computational efficiency, since all the fibers directions are \textit{a priori} included. \citet{Latorre2016} showed that the criterion originally chosen for excluding the compressed fibers, which is based on the mean direction~\citep{Gasser2006}, may produce nonphysical discontinuities in the stress–strain behavior. They proposed a new pre-integrated method, although limited to axially symmetric fiber distributions. \citet{Melnik2015, Li2018a} introduced a GST-based model in which the structure tensor is integrated accounting for the stretched fibers only, but since the computation is required at each deformation the efficiency of the original model is compromised. On the other hand, in the AI approach, also referred to as Continuous Fiber Dispersion (CFD)~\cite{Li2018}, the PDF is used to weigh the integration of the single fiber strain-energy over all the directions, allowing for a simple fiber exclusion criterion during the computation, and independent of the shape of the PDF. However, this comes with a compromise on the computational efficiency, since the integration must be performed multiple times during the analysis.

In an effort to increase the computational performance while preserving the advantages of the AI approach, an hybrid approach named Discrete Fiber Dispersion (DFD) model was introduced, based on a discrete integration over the spherical domain of the fibers~\cite{Li2018}. In this method, the choice of the integration scheme plays a fundamental role and can significantly affect the mechanical response. \citet{Li2018} originally proposed an integration method based on the discretization of the unit sphere with uniform triangular areas, each associated to a discrete fiber orientation with a weight computed by exactly integrating the PDF over the relative area. The method has been optimized by \citet{Rolf-Pissarczyk2021a} who developed an adaptive meshing to locally discretize the unit sphere surface with finer elements where the values of the PDF are concentrated. However, this method tends to increase the number of integration points for fiber distribution approaching the perfect alignment, while in such case only few representative directions would be sufficient. Moreover, there is little control on the number of integration points, since the refinement bisects spherical triangular elements until all the weights attain a value lower than a specified threshold.

In the light of the above, alternative quadrature schemes appear a valid choice for the integration due to their versatility for selecting the level of precision and the number of points~\cite{Itskov2016}. Notable examples include methods proposed by \citet{Bazant1986, Fliege1999, Heo2000}. There are mathematical requirements that need to be satisfied. To ensure poly-convexity of the strain-energy function~\cite{Ciarlet1988}, integration points should not be associated to negative coefficients~\cite{Ehret2010}. Moreover, the distribution of the points over the sphere should be as uniform as possible in order to maximize the integration efficiency, as well as to attain the invariance of the integral under general rotations of the integration points~\cite{Ehret2010, Skacel2015}. The number of points also plays a crucial role, since schemes with reduced points may induce anisotropic behaviors for isotropic materials, as pointed out by \citet{Ehret2010}. Lastly, since stress, strain, and related kinematic variables have identical values at diametrically opposite points on the sphere, centrally symmetric integration schemes are favored. They indeed allow the integral to be computed over only half of the sphere, effectively reducing the number of integrand evaluations by half~\cite{Alastrue2009}. Recently, \citet{Britt2023} proposed an approach based on the transformation of the spherical integral over the unit sphere into a linear integral over the distribution of the stretch of the fibers, which can be easily computed using a $n$-point Gauss quadrature.

According to the findings of \citet{Ehret2010}, who compared the performance of different integration schemes, quadrature schemes based on the construction of a spherical grid with polyhedral symmetry showed great potential in terms of computational efficiency and accuracy. In particular, the Lebedev rule, built exploiting octahedral symmetry~\cite{Lebedev1977,Lebedev1999}, demonstrated to perform the best out of the nine methods tested by \citet{Skacel2015}. However, only elementary loading cases were considered, and further analyses including complex loading configurations are required to investigate the convergence rate, as well as the ideal number of integration points for efficient Finite Element (FE) simulations. To this aim, in the present work, we address the numerical performance of the quadrature scheme with both elementary and advanced benchmark tests in large deformations. Specifically, we analyze the homogeneous uniaxial tension and simple shear tests, and the isotropic annular disk subjected to torsion of the internal boundary. The tests conducted confirm that the Lebedev quadrature exhibits rapid convergence to the exact solution, validating its suitability for high-fidelity mechanical simulations. The analyses were performed in the commercial FE software Abaqus~\citep{abaqus2018}. 

As an applicative example of the DFD model considered, we show a Z-plasty simulation, which has great relevance in skin mechanics. Z-plasty involves making Z-shaped incisions to release tension and improve the mobility of the skin, particularly in areas of scarring or contracture~\cite{Hove2001}. While the procedure is widely used in clinical practice, its application is largely guided by empirical experience rather than by a quantitative understanding of the underlying mechanics. If not correctly executed, these surgeries could potentially worsen the existing conditions, with the consequent need of further and more risky interventions. Therefore, a preoperative inspection is extremely important, and surgeons must assess key factors such as skin quality, underlying bone structure, vascular supply, extension of the defect, and direction of the natural skin tension. By combining this information the surgeon can determine the appropriate technique and plan the operation, as well as inform the patient about risks and possible outcomes~\cite{Baker2007}. In this regard, several surgical guidelines can be found in the literature~\cite{Rohrich1999, Humphreys2009, Hundeshagen2017}, some of which providing an algorithmic approach to determine the most suitable technique depending on factors such as the defect length, skin availability or skin tension~\cite{Leedy2005}. However, these guidelines all lack quantitative mechanical evidences.

Recent works of \citet{BuganzaTepole2014a} and \citet{Stowers2021} extensively analyzed similar surgical procedures, providing valuable mechanical insights that could help refine surgical practices, shifting from empirically driven decisions to more data-informed approaches. Following the same approach, in this work the DFD model is applied to simulate Z-plasty for various incision configurations and fiber orientations, providing novel insights into the mechanical behavior of the skin after the procedure. The simulation results can provide support for surgical practice, especially in terms of optimizing incision placement and orientation relative to the collagen fiber architecture of the skin. For example, the findings suggest that certain incision angles with respect to the mean collagen direction lead to reduced post-operative strain, thus improving healing outcomes. These results not only enhance the understanding of Z-plasty mechanics but also pave the way for more personalized and optimized surgical strategies.

The paper is organized as follows. In \cref{sec:mechanical_modeling} the theoretical framework of the DFD model is presented, including the derivation of the stress and stiffness tensors required for the FE implementation, and their specific formulation for incompressible plane stress problems. The Lebedev quadrature scheme is then introduced, illustrating the particular choice for the orientation of the integration points. Next, the convergence of the quadrature is assessed by means of numerical benchmark tests. In \cref{sec:Z-plasty_analysis} the Z-plasty is simulated for multiple geometrical configurations and orientations of the fibers. Results are provided in terms of key mechanical quantities relevant for the surgical practice. In \cref{sec:discussion} some considerations about the implemented model and the Z-plasty results are addressed, while in \cref{sec:conclusions} the conclusions are drawn.

\section{Mechanical modeling}\label{sec:mechanical_modeling}

\subsection{Description of general fiber distributions}\label{sec:fiber_dispersion}
\begin{figure}[t]
	\centering
	\includegraphics[width=13cm]{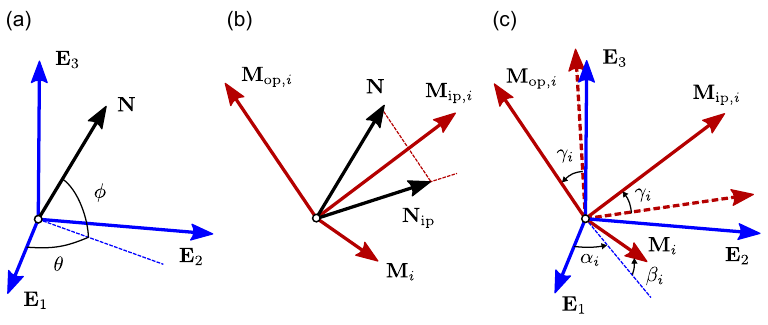}
	\caption{Schematic representation of the orientation of the unit vector $\N$ (in black), the global Cartesian basis $\{\E{1},\, \E{2},\, \E{3}\}$ (in blue) and the principal orthonormal basis $\{\Mi,\, \Mipi,\, \Mopi\}$ (in red) of the fiber family. (a) Unit vector $\mathbf{N}$ in the global frame; (b) unit vector $\mathbf{N}$ in the principal fiber family basis and its projection $\Nip$ onto the mean fiber plane $\{\Mi,\Mipi\}$; (c) rotation of the principal basis with respect to the global basis using the angles $\alpha_i$, $\beta_i$, $\gamma_i$. Dashed vectors represent $\Mipi$ and $\Mopi$ before the rotation about $\Mi$.}
	\label{fig:principal_fibers_frame}
\end{figure}

Let $\rho\parr{\N}=\rho\parr{\theta,\phi}$ be the Probability Density Function (PDF) defined over the unit sphere $\mathbb{S}=\{\parr{\theta,\phi}|\theta\in\pars{-\pi,\pi},\phi\in\pars{-\pi/2,\pi/2}\}$. The PDF provides the normalized collagen fiber density in the direction of the unit vector $\N$ in the reference configuration,
\begin{equation}
	\N\parr{\theta,\phi}=\cos\phi\cos\theta\E{1}+\cos\phi\sin\theta\E{2}+\sin\phi\E{3}\eend,
	\label{eq:fib_unit_vector_N}
\end{equation}
where $\{\E{1},\, \E{2},\, \E{3}\}$ defines the Cartesian unit vectors basis (\cref{fig:principal_fibers_frame}(a)), while $\theta$ and $\phi$ represent the azimuth and elevations angles, respectively. Since a fiber in the direction $\N$ is also represented by the opposite vector $-\N$, we require the PDF to satisfy the symmetry $\rho\parr{\N}=\rho\parr{-\N}$, or equivalently $\rho\parr{\theta,\phi}=\rho\parr{\theta+\pi,-\phi}$.

\begin{figure}[t]
	\centering
	\includegraphics[width=18cm]{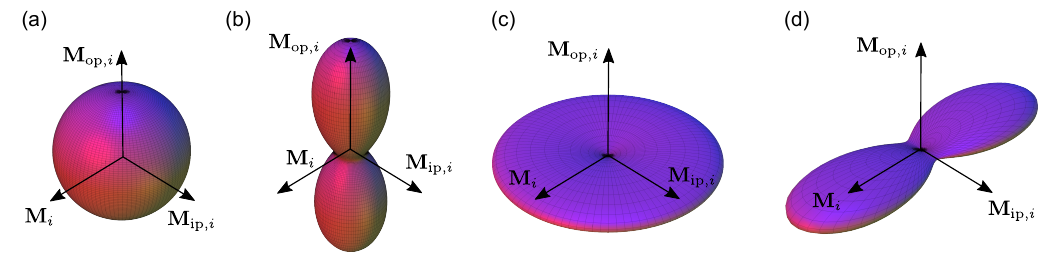}
	\caption{Polar representation of representative von Mises distributions. (a) Isotropic distribution, $a_i$, $b_i\rightarrow 0$; (b) axially symmetric distribution, $a_i=0$ and $b_i<0$; (c) quasi-planar isotropic distribution $a_i=0$, $b_i\rightarrow \infty$; (d) general distribution $a_i$, $b_i>0$.}
	\label{fig:repr_vM_distribs}
\end{figure}

We now assume the PDF to be a combination of a finite number $m$ of fiber families each described by an independent PDF $\rho_{i}\parr{\N}$ as
\begin{equation}
	\rho\parr{\N} = \sum_{i=1}^{m}\nu_i\rho_{i}\parr{\N}\eend,
	\label{eq:fib_global_fiber_distribution}
\end{equation}
where $\nu_i$ represents the normalized volume fraction of the $i$-th fiber family, with $\Sigma\nu_i=1$, and each $\rho_{i}\parr{\N}$ must satisfy the normalization condition over the unit sphere $\mathbb{S}$
\begin{equation}
	\dfrac{1}{4\pi}\int\limits_{\mathbb{S}}\rho_{i}(\N)\de S = \dfrac{1}{4\pi}\int_{-\pi}^{\pi}\int_{-\pi/2}^{\pi/2}\rho_{i}\parr{\theta,\phi}\cos\phi\de\theta\de\phi = 1\eend.
	\label{eq:fib_distr_normalization}
\end{equation}
To account for general orientations of each fiber family in the Euclidean space we then introduce the principal orthonormal unit vectors basis $\{\Mi,\, \Mipi,\, \Mopi\}$ (\cref{fig:principal_fibers_frame}(b)), in which $\Mi$ represents the preferential direction of the family, $\{\Mi,\Mipi\}$ defines its mean plane, and $\Mopi$ is the out-of-plane normal. The rotation of this basis relative to the global basis $\{\E{k}\}_{k=1,2,3}$ is described by the triplet of angles $\alpha_{i}$, $\beta_{i}$, $\gamma_{i}$, where the first two denote the azimuthal and elevation angle of the mean fiber direction, $\Mi=\mathbf{N}(\alpha_i,\beta_i)$, while the last one represents the rotation of the basis about $\Mi$ (\cref{fig:principal_fibers_frame}(c)). Accordingly, based on experimental evidences on biological tissues~\cite{Niestrawska2016,Alberini2024}, the PDF $\rho_{i}$ is assumed to be decoupled into two independent univariate PDFs as
\begin{equation}
	\rho_{i}\parr{\N} = \rho_{\ip,i}\parr{\N}\rho_{\op,i}\parr{\N}\eend,
	\label{eq:fib_bivariate_fiber_distribution}
\end{equation}
in which $\rho_{\ip,i}\parr{\N}$ and $\rho_{\op,i}\parr{\N}$ describe, respectively, the in-plane and out-of-plane distributions with respect to the basis $\{\Mi,\, \Mipi,\, \Mopi\}$.
Following the work of \citet{Holzapfel2015}, we specialize the two functions employing two $\pi$-periodic von Mises distributions of the form
\begin{align}
	\rho_{\ip,i}(\mathbf{N}) &= \dfrac{\exp[a_i(2(\Nip\cdot\Mi/\lvert\Nip\rvert)^2-1)]}{I_0(a_i)}\eend, \label{eq:fib_vonMises_ip}\\
	\rho_{\op,i}(\mathbf{N}) &= 2\sqrt{\dfrac{2b_i}{\pi}}\dfrac{\exp[-2b_i(\mathbf{N}\cdot\Mopi)^2]}{\erf(\sqrt{2b_i})}\eend, \label{eq:fib_vonMises_op}
\end{align}
where the vector $\Nip$ represent the projection of $\N$ onto the plane $\{\Mi,\Mipi\}$. The functions $I_0(\bullet)$ and $\erf(\bullet)$ are the modified Bessel function of the first kind of order zero and the error function of $(\bullet)$, respectively, which ensure the normalization condition \eqref{eq:fib_distr_normalization}. The constants $a_i$ and $b_i$ are parameters that define the in-plane and out-of-plane concentrations, respectively. Thus, using \cref{eq:fib_global_fiber_distribution,eq:fib_bivariate_fiber_distribution} each fiber family is fully described by the set of parameters $\{a_i,\, b_i,\, \alpha_i,\, \beta_i,\, \gamma_i,\, \nu_i \}$. The limit case of $a_i$, $b_i\rightarrow\infty$ corresponds to perfect fiber alignment along $\Mi$, while for $a_i$, $b_i\rightarrow 0$ an isotropic distribution is obtained (\cref{fig:repr_vM_distribs}(a)). Negative values are also admissible, but describe a fiber distribution no longer concentrated in the direction $\Mi$. For instance, $a_i=0$ and $b_i<0$ represent an axially symmetric fiber distribution with preferential direction along $\Mopi$ (\cref{fig:repr_vM_distribs}(b)). The particular choice for the distributions in \cref{eq:fib_vonMises_ip,eq:fib_vonMises_op} verifies symmetry requirement $\rho_{i}\parr{\N}=\rho_{i}\parr{-\N}$, and has the additional in-plane and out-of-plane symmetries about $\Mi$.

\subsection{Constitutive model}\label{sec:model}
Let $\F=\partial\x\parr{\X}/\partial\X$ be the deformation gradient describing the deformation field in the neighborhood of a material point moving from the position $\X$ in the reference configuration to a position $\x$ in the current configuration. The change in length and direction of a fiber in the direction of the unit vector $\N$ in the reference configuration can then be expressed as $\F\N=\lambda\n$, where $\lambda$ is the stretch ratio and $\n$ is the unit vector of the new fiber direction in the current configuration. Accordingly, the square of the fiber stretch is expressed by the invariant $I_{4}=\lambda^2=\C:\N\otimes\N$, with $\C=\F^{\mathrm{T}}\F$ denoting the right Cauchy-Green tensor.

Following the DFD approach, we describe the mechanical behavior of a single fiber family by approximating the overall strain-energy function of the continuous fiber dispersion as~\cite{Li2018}
\begin{equation}
	\Psi_{\mathrm{f},i} = \frac{1}{4\pi}\int_{\mathbb{S}}\rho_{i}(\N)\Psi_{\mathrm{sf}}(I_{4})\de S \approx \sum_{j=1}^{n}w_{j}\rho_{i}(\N_{j})\Psi_{\mathrm{sf}}(\Ifour{j})\eend,
	\label{eq:Psi_DFD}
\end{equation}
where $\N_{j}$ and $w_{j}$, $j=1,\dots,n$, are the integration points and weights defined by the specific integration scheme over the unit sphere $\mathbb{S}$, respectively, $\Psi_{\mathrm{sf}}$ is the single-fiber strain-energy function, and $\Ifour{j}=\C:\N_{j}\otimes\N_{j}$ is the squared stretch of a fiber in the direction of $\N_{j}$. In general, the number $n$ of integration points can be different for each fiber family. To exclude the negligible contribution of the compressed fibers, the single-fiber strain-energy function $\Psi_{\mathrm{sf}}(\Ifour{j})$ is defined as
\begin{equation}
	\Psi_{\mathrm{sf}}(\Ifour{j}) = \begin{cases}
		f\parr{\Ifour{j}} & \Ifour{j}\geq1,\\
		0 & \Ifour{j}<1,
	\end{cases}
	\label{eq:Psi_single_fiber}
\end{equation}
adopting the exponential function proposed by \citet{Holzapfel2000}, i.e.
\begin{equation}
	f(\Ifour{j}) = \dfrac{c_{1}}{2 c_{2}}\left\lbrace \exp\left[c_{2}(\Ifour{j}-1)^{2}\right]-1\right\rbrace\eend.
	\label{eq:exp_function_Holzapfel}
\end{equation}
in which the two constants $c_1$ and $c_2$ represents the stiffness-like parameter and the stiffening parameter of the fibers, respectively.

The mechanical behavior of the isotropic ground matrix embedding the collagen fibers is described using the neo-Hookean model, whose strain-energy function reads
\begin{equation}
	\Psi_{\mathrm{g}} = \dfrac{\mu}{2}\parr{I_{1}-3}\eend,
	\label{eq:Psi_neo-Hookean}
\end{equation}
where $I_{1}=\C:\I$ is the first invariant of $\C$, $\I$ is the second-order identity tensor, and $\mu$ is the initial shear modulus.

Assuming material incompressibility, a common assumption in models of soft biological materials~\cite{Pissarenko2020}, the total strain-energy function $\Psi$ is obtained by additively combining the energy contributions of the ground matrix and of the fiber families weighed with their respective volume fractions $\nu_i$, i.e.
\begin{equation}
	\Psi = \Psi_{\mathrm{g}}+\sum_{i=1}^{m}\nu_i\Psi_{\mathrm{f},i} - p(J-1)\eend,
	\label{eq:Psi_tot}
\end{equation}
where the last term depending on the volume ratio $J=\det\parr{\F}$ is a penalty function introduced to enforce perfect incompressibility ($J=1$), in which the Lagrange multiplier $p$ assumes the meaning of an unknown hydrostatic pressure.

\subsection{Stress and stiffness tensors for incompressible plane stress formulations}\label{sec:tensors}
The implementation of the model \eqref{eq:Psi_tot} into a FE code requires the stress and stiffness tensors to be explicitly computed.

The second Piola-Kirchhoff stress tensor $\mathbf{S}$ is obtained by differentiation of \cref{eq:Psi_tot} with respect to $\frac{1}{2}\C$. Since both $J$ and $p$ are dependent on $\C$, the tensor reads~\cite{Kiendl2015}
\begin{equation}
	\mathbf{S} = 2\frac{\partial\Psi}{\partial\C} = 2\left[\Psi_{\mathrm{g}}^{'}(I_{1})\I+\sum_{i=1}^{m}\nu_i\sum_{j=1}^{n}\varrho_{i,j}\Psi_{\mathrm{sf}}^{'}(\Ifour{j})\N_{j}\otimes\N_{j}\right] - 2\dfrac{\partial p}{\partial\C}(J-1)-pJ\C^{-1}\eend,
	\label{eq:stress_PKII}
\end{equation}
with $\Psi^{'}_{\lozenge}({\scriptstyle {\blacklozenge}})$ denoting the first derivative of $\Psi_{\lozenge}$ with respect to ${\scriptstyle {\blacklozenge}}$, and $\varrho_{i,j}=w_{j}\rho_{i}(\N_{j})$ is introduced for shorter notation. Note that the term containing the derivative of $p$, even thought it would vanish for $J=1$, is kept to ensure the full derivation of the spherical component of the stiffness tensor~\cite{Kiendl2015}. By a Piola transformation on $\mathbf{S}$ the Cauchy stress tensor $\bm{\sigma}$ can be obtained as
\begin{equation}
	\bm{\sigma} = J^{-1}\F\mathbf{S}\F^{\mathrm{T}} = 2J^{-1}\left[\Psi_{\mathrm{g}}^{'}(I_{1})\b+\sum_{i=1}^{m}\nu_i\sum_{j=1}^{n}\varrho_{i,j}\Psi_{\mathrm{sf}}^{'}(\Ifour{j})\n_{j}\otimes\n_{j}\right]
	-2J^{-1}\b\dfrac{\partial p}{\partial\b}(J-1)-p\I\eend,
	\label{eq:stress_Cauchy}
\end{equation}
where $\b=\F\F^{T}$ is the left Cauchy-green tensor, and $\n_{j}=\F\N_{j}$.

Further differentiation of \cref{eq:stress_PKII} with respect to $\frac{1}{2}\C$ gives the tangent elasticity tensor $\mathbb{C}$ in Lagrangian description
\begin{flalign}
	\begin{split}
		\mathbb{C}=2\frac{\partial\mathbf{S}}{\partial\C} =& 4\left[\Psi_{\mathrm{g}}^{''}(I_{1})\I\otimes\I+\sum_{i=1}^{m}\nu_i\sum_{j=1}^{n}\varrho_{i,j}\Psi_{\mathrm{sf}}^{''}(\Ifour{j})\N_{j}\otimes\N_{j}\otimes\N_{j}\otimes\N_{j}\right] + \\
		&-4\dfrac{\partial^{2} p}{\partial\C\partial\C}(J-1)-2J\left(\dfrac{\partial p}{\partial\C}\otimes\C^{-1}+\C^{-1}\otimes\dfrac{\partial p}{\partial\C}\right) + \\
		&+2pJ\C^{-1}\odot\C^{-1}-pJ\C^{-1}\otimes\C^{-1}\eend,
		\label{eq:stiff_PKII}
	\end{split}
\end{flalign}
with $\Psi^{''}_{\lozenge}({\scriptstyle {\blacklozenge}})$ denoting the second derivative of $\Psi_{\lozenge}$ with respect to ${\scriptstyle {\blacklozenge}}$, and the operator $\odot$ defining the symmetric dyadic product between two second order tensors, $(\circ\odot\bullet)_{ijkl}=1/2((\circ)_{ik}(\bullet)_{jl}+(\circ)_{il}(\bullet)_{jk})$. Similarly to \cref{eq:stress_Cauchy}, a push-forward transformation of the fourth-order tensor $\mathbb{C}$ gives the tangent elasticity tensor $\stiffc$ in spatial description~\cite{Holzapfel2000book}
\begin{flalign}
	\begin{split}
		\stiffc = J^{-1}\chi_{\ast}(\mathbb{C}) =& J^{-1}4\left[\Psi_{\mathrm{g}}^{''}(I_{1})\b\otimes\b+\sum_{i=1}^{m}\nu_i\sum_{j=1}^{n}\varrho_{i,j}\Psi_{\mathrm{sf}}^{''}(\Ifour{j})\n_{j}\otimes\n_{j}\otimes\n_{j}\otimes\n_{j}\right] + \\
		&-4J^{-1}\b\dfrac{\partial^{2} p}{\partial\b\partial\b}\b(J-1)-2\left(\b\dfrac{\partial p}{\partial\b}\otimes\I+\I\otimes\dfrac{\partial p}{\partial\b}\b\right) + \\
		&+2p\I\odot\I-p\I\otimes\I\eend.
		\label{eq:stiff_Cauchy}
	\end{split}
\end{flalign}

For general three-dimensional problems the pressure $p$ is unknown and it is treated as an independent solution variable commonly solved with a hybrid FE formulation~\cite{Crisfield1996book}. For plane stress problems, assuming the basis vectors $\E{1},\, \E{2}$ laying in the mean plane, the explicit expression of the pressure $p$ can be derived from \cref{eq:stress_Cauchy} imposing the constraints $\sigma_{33}=\bm{\sigma}:\parr{\E{3}\otimes\E{3}}=0$ and $J=1$, namely
\begin{equation}
	p = \tilde{\bm{\sigma}}:\parr{\E{3}\otimes\E{3}} = 2J^{-1}\left[\Psi_{\mathrm{g}}^{'}(I_{1})(\b)_{33}+\sum_{i=1}^{m}\nu_i\sum_{j=1}^{n}\varrho_{i,j}\Psi_{\mathrm{sf}}^{'}(\Ifour{j})(\n_{j}\otimes\n_{j})_{33}\right]\eend,
	\label{eq:pressure}
\end{equation}
with $\tilde{\bm{\sigma}}=\bm{\sigma} + p\bm{I}$ denoting the deviatoric component of the Cauchy stress. The derivatives of $p$ in \cref{eq:stiff_Cauchy} can then be explicitly computed from \cref{eq:pressure} giving
\begin{equation}
	\b\dfrac{\partial p}{\partial\b} = \dfrac{\partial p}{\partial\b}\b = \dfrac{1}{2}\parr{\E{3}\otimes\E{3}}:\tilde{\stiffc}- \dfrac{1}{2}p\I+p\E{3}\otimes\E{3}\eend,
	\label{eq:bdpdb}
\end{equation}
where $\tilde{\stiffc}$ is the deviatoric component of the spatial stiffness tensor, represented by the terms within the square brackets in \cref{eq:stiff_Cauchy}. The second derivative of $p$ in \cref{eq:stiff_Cauchy} is not computed since the relevant term vanishes for $J=1$.

\subsection{Implementation into Abaqus FE software}
The model presented in \cref{sec:model} and \cref{sec:tensors} has been implemented in the commercial FE software Abaqus through the user-defined subroutine UMAT.

According to the implicit solver formulation (Abaqus/Standard) for plane stress elements, the tensors to be defined are the Cauchy stress, as provided in \cref{eq:stress_Cauchy}, and the tangent stiffness tensor $\overset{\circ}{\stiffc}$ relative to the Jaumann rate of the Kirchhoff stress $\overset{\circ}{\bm{\tau}}=\stiffc:\mathbf{d}+\mathbf{d}\bm{\tau}+\bm{\tau}\mathbf{d}=\overset{\circ}{\stiffc}:\mathbf{d}$~\cite{Prot2007,Palizi2020}, i.e.
\begin{equation}
	\overset{\circ}{\stiffc} = \stiffc+\bm{\tau}\odot\I+\I\odot\bm{\tau}\eend,
	\label{eq:stiff_Jaumann}
\end{equation}
where $\mathbf{d}=\symm(\overset{.}{\F}\F^{-1})$ is the rate of deformation tensor, and $\bm{\tau}=J\bm{\sigma}$ is the Kirchhoff stress tensor.

Since only the planar strain components are used in the plane stress formulation, the transverse deformation $(\C)_{33}$ is statically condensed and eliminated from the problem. Accordingly, the coefficients of the statically condensed stiffness tensor $\hat{\stiffc}$ are obtained as~\cite{Kiendl2015}

\begin{equation}
	\parr{\hat{\stiffc}}_{ijkl} = (\overset{\circ}{\stiffc})_{ijkl}-\dfrac{(\overset{\circ}{\stiffc})_{ij33}(\overset{\circ}{\stiffc})_{33kl}}{(\overset{\circ}{\stiffc})_{3333}}\eend.
	\label{eq:stiff_Jaumann_condens}
\end{equation}

Additionally, the model has been implemented for perfectly incompressible 3D problems in hybrid formulation. For this case, the Cauchy stress tensor and the full stiffness tensor in \cref{eq:stiff_Jaumann} have been defined into the subroutine using only their deviatoric components $\tilde{\bm{\sigma}}$ and $\tilde{\stiffc}$, respectively. The spherical components are not required, since they are computed by the Abaqus/Standard solver as part of the problem using the hydrostatic pressure variable $p$.

The special functions $I_0(\bullet)$ and $\erf(\bullet)$ needed to compute the distributions in \cref{eq:fib_vonMises_ip,eq:fib_vonMises_op} are incorporated in the UMAT subroutine using the Fortran codes provided by \citet{Zhang1996}.

\subsection{Model integration scheme}\label{sec:Lebedev_integration}
\begin{figure}[ht]
	\centering
	\includegraphics[width=9cm]{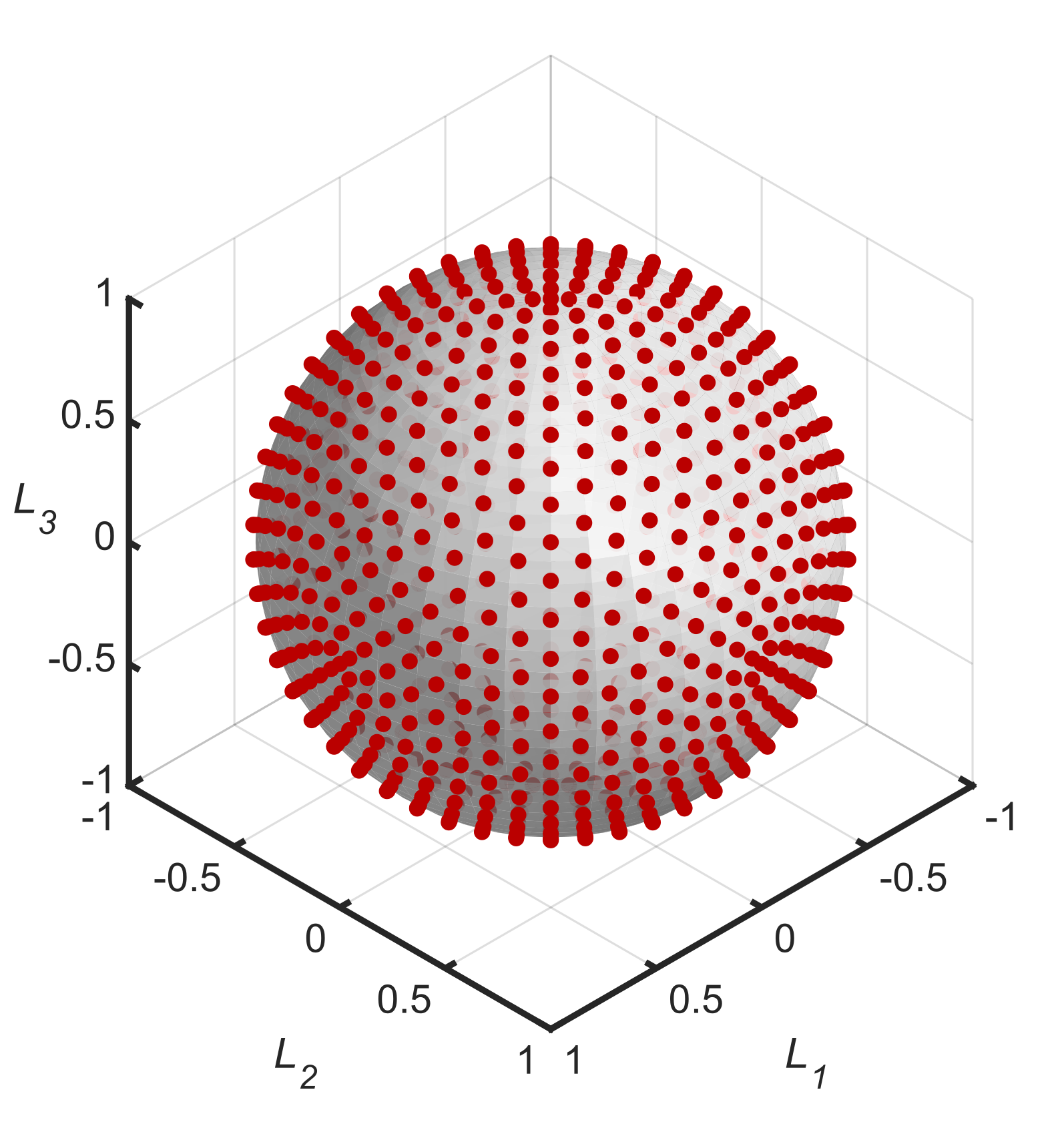}
	\caption{Representative spherical grid of $n=2\times487$ Lebedev integration points in the basis $\{\mathbf{L}_{k}\}_{k=1,2,3}$.}
	\label{fig:Lebedev_points}
\end{figure}

For the numerical integration of \cref{eq:Psi_DFD} we adopted the Lebedev quadrature rule. The integration points $\N_{j}$ of this scheme are placed on the surface of the unit sphere forming a grid with octahedral rotational and inversion symmetries about a given orthonormal basis $\{\mathbf{L}_{k}\}_{k=1,2,3}$, while the relative weights $w_j$, as well as the number of points $n$, are determined by enforcing the exact integration of all the spherical harmonics up to a specific degree. Thus, the order of the integration is defined by the degree of the exactly integrated spherical harmonic. To date, Lebedev quadratures have been solved for degrees up to $131$, corresponding to $n=5810$ integration points~\cite{Lebedev1999}. In contrast to methods based on the Cartesian product of one-dimensional integrations which cause points to densify near the poles, such as the Gaussian quadrature, Lebedev quadratures yield an even distribution of points across the sphere, as shown in \cref{fig:Lebedev_points} for $n=974$. As a result, the number of points per unit area is approximately the same over the unit sphere, thereby enhancing the computational efficiency~\cite{Beentjes2015,Skacel2015}. Another important aspect of this approach is that the weights $w_j$ are always positive, ensuring the poly-convexity of the solution. In fact, if any weights were negative, the integration could lead to unrealistic results, like negative stresses, when the deformations is all concentrated along a point with negative weight.

For functions different from a spherical harmonic, the Lebedev quadrature provides an approximation of the exact integral, with accuracy improving as the order of the integration increases. Moreover, as for all quadratures, the integral should ideally remain invariant under rigid rotations of the set of integration points relative to the integrand function. With reference to \cref{eq:Psi_DFD}, this implies that any rotation of the basis $\{\mathbf{L}_{k}\}_{k=1,2,3}$ relative to the global basis $\{\E{k}\}_{k=1,2,3}$ yields theoretically the same result. Therefore, without loss of generality, we align $\{\mathbf{L}_{k}\}_{k=1,2,3}$ with the principal fiber basis $\{\Mi,\, \Mipi,\, \Mopi\}$, so that the integration points $\N_{j}$ and the distribution $\rho_{i}\parr{\N}$ share the same symmetries. This makes the mechanical response perfectly symmetrical for symmetric deformations about the principal fiber basis, thus ensuring the consistency of the results. Moreover, this guarantees the existence of a point $\N_{j}$ for every vector of the principal fiber basis. Such a property is useful for nearly perfectly aligned fiber distributions, where the mechanical response is dominated by the integration point $\N_{j}$ aligned with the mean fiber direction, which can be either $\Mi$ for $a$, $b\rightarrow\infty$, or $\Mopi$ for $a=0$, $b\rightarrow-\infty$.

By exploiting the symmetry $\rho_{i}\parr{\N}=\rho_{i}\parr{-\N}$ discussed in \cref{sec:fiber_dispersion}, we can reduce the computational cost by restricting the integration to only one half of the spherical domain $\mathbb{S}$. The number of integration points can be further reduced by eliminating the directions $\N_{j}$ associated to a negligible fiber density, e.g. $\varrho_{i,j}=w_{j}\rho_{i}(\N_{j})<10^{-6}$. In this way, as the fiber concentration increases, the number of significant directions $\N_{j}$ reduces, degenerating to only one direction for a perfectly aligned fiber distribution.

\section{Benchmark tests for convergence assessment}\label{sec:convergence_assessment}
The number $n$ of integration points must be carefully selected to balance computational efficiency and accurate results. To this end, we analyzed the convergence rate of representative problems to the reference solution in relation to the integration order. Specifically, we considered two simple cases of a uniaxial and a simple shear test using a single fiber family with given distribution parameters, and a benchmark torsion test on an annular disk with planar isotropic fiber distribution. For each case we considered integration orders defined by $n=55$, $97$, $151$, $175$, $217$, $295$, $385$, $487$, $601$, $727$ points over one hemisphere of the unit sphere. For comparison, in the uniaxial and and simple shear tests we also included the results of the DFD model computed using the adaptive integration rule proposed by \citet{Rolf-Pissarczyk2021a}. The points were generated using the \textsc{Matlab} script provided by the authors. Since the number of points cannot arbitrarily defined, we obtained three sets of integration points with $n=100$, $292$, $808$, by varying the threshold value for $\varrho_{i,j}$.

\subsection{Uniaxial tension}\label{sec:convergence_uniax}
\begin{figure}[t]
	\centering
	\includegraphics[width=18cm]{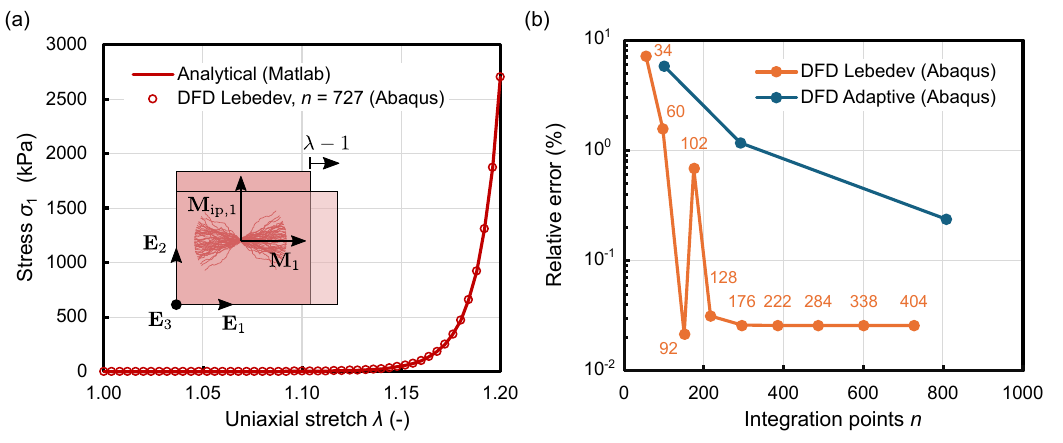}
	\caption{DFD model convergence assessment for a unit square element in uniaxial tension. (a) Cauchy stress $\sigma_{1}$ versus the stretch $\lambda$ from the CFD model solution in \cref{eq:sigma1_analytical} (solid line) and the representative DFD model solution with $n=727$ (circles) for the set of mechanical parameters $\mu=1.142\ \text{kPa}$, $c_1=1.9074\ \text{kPa}$, $c_2=43.6$, and fiber distribution parameters $a_1=2.06$, $b_1=12.36$, $\alpha_1=\beta_1=\gamma_1=0^{\circ}$. A sketch of the uniaxial test is also reported, showing a schematic representation the fiber distribution orientation in relation to the unit square element. (b) Convergence rate of the DFD model implemented in Abaqus relative to the reference CFD solution computed in \textsc{Matlab} for the Lebedev quadrature rule (orange dots) and the Adaptive integration scheme of \citet{Rolf-Pissarczyk2021a} (blue dots). Orange numbers indicate the effective number of integration points used, satisfying the condition $\varrho_{i,j}=w_{j}\rho_{i}(\N_{j})\geq10^{-6}$.}
	\label{fig:dfd_uniaxial_test}
\end{figure}

In this case, we considered a unit square plane stress element (Abaqus CPS4 element) extended along the $\E{1}$ direction by a stretch $\lambda=1.2$. The material parameters were taken from \citet{Alberini2024a}, with mechanical constants $\mu=1.142\ \text{kPa}$, $c_1=1.9074\ \text{kPa}$, $c_2=43.6$, and fiber concentrations $a_1=2.06$, $b_1=12.36$, obtained by averaging the microstructural parameters of human skin (Ref~\cite{Alberini2024a} Table 2, Fiber family 1). The fibers were oriented in the direction of the stretch, such that $\mathbf{M}_1=\E{1}$ ($\alpha_1=\beta_1=\gamma_1=0^{\circ}$). We used as reference solution the CFD model of \citet{Li2016b}, which provides the components of the Cauchy stress tensor derived from the exact integration of \cref{eq:Psi_DFD}. For a uniaxial tests along the $\E{1}$ direction, the analytical expression of the Cauchy stress $\sigma_{1}$, accounting for the non-symmetrical in-plane and out-of-plane fiber dispersions as defined in \cref{eq:fib_bivariate_fiber_distribution}, reads
\begin{equation}
	\sigma_{1} = \parr{\mu+Y_1}\lambda^{2}-\sqrt{\parr{\mu+Y_2}\parr{\mu+Y_3}}\lambda^{-1}\eend,
	\label{eq:sigma1_analytical}
\end{equation}
with
\begin{align} 
	Y_1 =& \dfrac{1}{\pi}\int_{\Sigma}\rho\parr{\theta,\phi}\Psi^{'}_{\mathrm{sf}}(I_{4}\parr{\theta,\phi})\cos^3\phi\cos^2\theta\ \de\theta\de\phi\eend, \label{eq:sigma1_Y1}\\
	Y_2 =& \dfrac{1}{\pi}\int_{\Sigma}\rho\parr{\theta,\phi}\Psi^{'}_{\mathrm{sf}}(I_{4}\parr{\theta,\phi})\sin^2\phi\cos\phi\ \de\theta\de\phi\eend, \label{eq:sigma1_Y2}\\
	Y_3 =& \dfrac{1}{\pi}\int_{\Sigma}\rho\parr{\theta,\phi}\Psi^{'}_{\mathrm{sf}}(I_{4}\parr{\theta,\phi})\cos^3\phi\sin^2\theta\ \de\theta\de\phi\eend, \label{eq:sigma1_Y3}
\end{align}
where $\Sigma=\left\{\theta\in\pars{-\pi,\pi},\phi\in\pars{-\pi/2,\pi/2}|I_{4}\parr{\theta,\phi}>1\right\}$. The deformation-dependent coefficients given in \cref{eq:sigma1_Y1,eq:sigma1_Y2,eq:sigma1_Y3} are solved numerically using a \textsc{Matlab} script with the built-in function \texttt{integral2}. As shown in \cref{fig:dfd_uniaxial_test}(a) for a representative integration with $n=727$, the result of the DFD model is in perfect agreement with the CFD model solution.

The convergence rates of the DFD model integrated using the Lebedev quadrature and the adaptive integration rule \citet{Rolf-Pissarczyk2021a} are reported in \cref{fig:dfd_uniaxial_test}(b). For the Lebedev quadrature the error rapidly decreases from about $7.1\%$ for $n=55$ to $0.03\%$ for $n=217$, and then it settles at $0.025\%$ for higher integration orders. The adaptive integration also reduces the error with $n$, but at a slower convergence rate. Notably, as shown in \cref{fig:dfd_uniaxial_test}(b), the effective number of points used, by excluding the direction $\N_{j}$ with negligible influence, is almost reduced by half for all the quadratures considered.

\subsection{Simple shear}\label{sec:convergence_shear}
\begin{figure}[t]
	\centering
	\includegraphics[width=18cm]{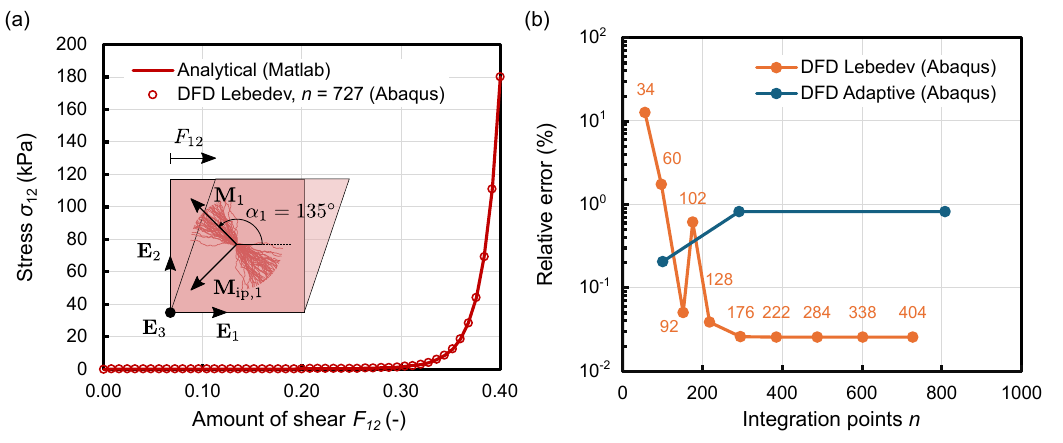}
	\caption{DFD model convergence assessment for a unit square element in simple shear. (a) Stress $\sigma_{12}$ versus the amount of shear $F_{12}$ plots of the CFD model solution in \cref{eq:sigma12_analytical} (solid line) and the representative DFD model solution with $n=727$ (circles) for the set of mechanical parameters $\mu=1.142\ \text{kPa}$, $c_1=1.9074\ \text{kPa}$, $c_2=43.6$, and fiber distribution parameters $a_1=2.06$, $b_1=12.36$, $\alpha_1=135^{\circ}$, $\beta_1=\gamma_1=0^{\circ}$. A sketch of the shear test is also reported, showing a schematic representation the fiber distribution orientation in relation to the unit square element. (b) Convergence rate of the DFD model implemented in Abaqus relative to the reference CFD model solution computed in \textsc{Matlab} for the Lebedev quadrature rule (orange dots) and the Adaptive integration scheme of \citet{Rolf-Pissarczyk2021a} (blue dots). Orange numbers indicate the effective number of integration points used, satisfying the condition $\varrho_{i,j}=w_{j}\rho_{i}(\N_{j})\geq10^{-6}$.}
	\label{fig:dfd_shear_test}
\end{figure}

In this case, we applied a displacement on the top edge of a unit square plane stress element (Abaqus CPS4 element) in the $\E{1}$ direction for an amount of shear of $F_{12}=0.4$. The material parameters are the same used in the uniaxial test in \cref{sec:convergence_uniax}. To enhance the exclusion of the fibers in compression, we oriented the fiber family at $\alpha_1=135^{\circ}$, while keeping $\beta_1=\gamma_1=0^{\circ}$. As previously, we used the CFD model as reference solution, for which the analytical expression of the Cauchy shear stress component $\sigma_{12}$ is given by
\begin{equation}
	\sigma_{12} = \parr{\mu+Y_4}F_{12}+Y_3\eend,
	\label{eq:sigma12_analytical}
\end{equation}
with
\begin{equation} 
	Y_4 = \dfrac{1}{\pi}\int_{\Sigma}\rho\parr{\theta,\phi}\Psi^{'}_{\mathrm{sf}}(I_{4}\parr{\theta,\phi})\cos^3\phi\sin\theta\cos\theta\ \de\theta\de\phi\eend,
	\label{eq:sigma12_Y4}
\end{equation}
where, again, the coefficients $Y_3$ from \cref{eq:sigma1_Y3} and $Y_4$ are computed numerically over the domain $\Sigma=\left\{\theta\in\pars{-\pi,\pi},\phi\in\pars{-\pi/2,\pi/2}|I_{4}\parr{\theta,\phi}>1\right\}$ using a \textsc{Matlab} script with the built-in function \texttt{integral2}. The representative result for $n=727$, and the reference CFD model solution are shown in \cref{fig:dfd_shear_test}(a).

In \cref{fig:dfd_shear_test}(b) are reported the convergence rates of the DFD model integrated using the Lebedev quadrature and the adaptive integration rule. As for the uniaxial case, the Lebedev quadrature rapidly converges, settling at an error of $0.026\%$ after $n=217$. For the adaptive integration the error initially increases from $ 0.21\%$ for $n=100$ to $0.82\%$ for $n=292$, but then it stabilizes. The effective numbers of integration points shown in \cref{fig:dfd_shear_test}(b) are the same as in the uniaxial tension test, since the same fiber distribution was used.

\subsection{Torsion of isotropic annular disk}
\label{sec:annular disk}
Beside the convergence rate to the exact solution, a crucial aspect of quadrature schemes, as pointed out by \citet{Ehret2010}, concerns the invariance of the integration upon rigid rotations of the points $\N_{j}$ about the integrand function. In fact, for quadratures with a poor number of integration points, the rotation of the basis $\{\mathbf{L}_{k}\}_{k=1,2,3}$ relative to the integrand function influences the result of the integration, thereby altering the mechanical response, which could become anisotropic for isotropic materials~\cite{Ehret2010, Skacel2015}. Conversely, as the number of integration points increases, the quadrature should return the same result regardless the orientation of $\{\mathbf{L}_{k}\}_{k=1,2,3}$.

To examine the rotational invariance with respect to the number of points $n$, one can analyze rotations of either the basis $\{\mathbf{L}_{k}\}_{k=1,2,3}$ or the integrand function while keeping the basis fixed. 
Since in \cref{sec:Lebedev_integration} we restricted $\{\mathbf{L}_{k}\}_{k=1,2,3}$ to be coaxial with the principal fiber basis $\{\Mi,\, \Mipi,\, \Mopi\}$, we then decided to analyze an axisymmetric problem in which a quasi-planar isotropic fiber distribution remains fixed throughout the whole domain $\Omega$, while the integrand function of \cref{eq:Psi_DFD} rotates with the local cylindrical basis $\{\E{r},\E{\vartheta},\E{3}\}$ with no dependency on $\vartheta$.
Specifically, following \citet{Ehret2010}, we considered an annular disk with inner radius $R_{\mathrm{in}}=4\ \text{mm}$, outer radius $R_{\mathrm{out}}=10\ \text{mm}$, and thickness $t=0.5\ \text{mm}$. The disk was discretized with 4-node plane stress elements (Abaqus CPS4 element), using $360$ and $49$ elements in the circumferential and radial direction, respectively. A counter-clockwise rotation of $\varepsilon=30^{\circ}$ was applied to the inner boundary $\partial\Omega_{\mathrm{in}}$ while keeping its radial displacements fixed, as well as all the displacements of the outer boundary $\partial\Omega_{\mathrm{out}}$, constrained. Mechanical parameters were the same used in the previous tests, using the concentration parameters $a_1=0$ and $b_1=12.36$ to represent a quasi-planar isotropic fiber dispersion, similar to that shown in \cref{fig:repr_vM_distribs}(c). Since the integration points remain fixed, the choice for the fiber distribution is arbitrary. We then used $\alpha_1=\beta_1=\gamma_1=0^{\circ}$, such that $\{\mathbf{L}_{k}\}_{k=1,2,3}\equiv\{\E{k}\}_{k=1,2,3}$.
\begin{figure}[t]
	\centering
	\includegraphics[width=7cm]{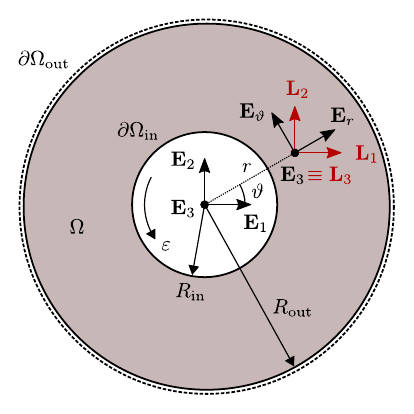}
	\caption{Torsion of an annular disk with respect to the vector $\E{3}$. The outer boundary $\partial\Omega_{\mathrm{out}}$ is constrained in the radial and tangential directions, indicated with a dashed line, while the inner boundary $\partial\Omega_{\mathrm{in}}$ is left-rotated by $\varepsilon=30^{\circ}$ about $\E{3}$, keeping fixed the radial displacement. The basis of the integration points $\{\mathbf{L}_{k}\}_{k=1,2,3}\equiv\{\Mi,\, \Mipi,\, \Mopi\}$ is aligned with $\{\E{k}\}_{k=1,2,3}$ over the domain $\Omega$. Due to the quasi-planar isotropic distribution chosen for the fibers, $a_1=0$, $b_1=12.36$, the integrand function in \cref{eq:Psi_DFD} evaluated with respect to the local cylindrical basis $\{\E{r},\E{\vartheta},\E{3}\}$ is independent on the angle $\vartheta$.}
	\label{fig:dfd_annular_disk_scheme}
\end{figure}

\begin{figure}[t]
	\vspace{-1cm}
	\centering
	\includegraphics[width=18cm]{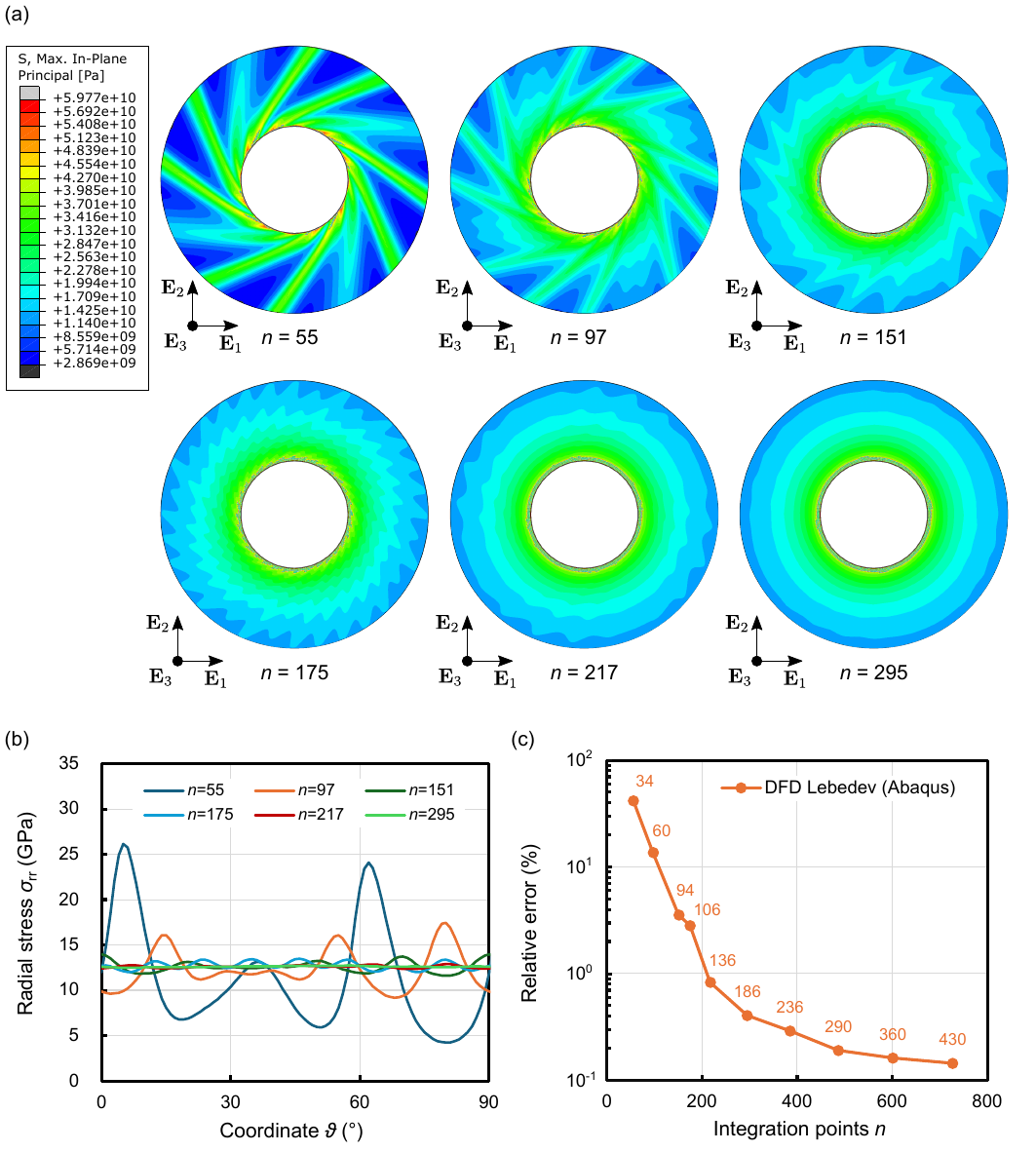}
	\caption{Rotational invariance assessment of the DFD model for an annular disk with mechanical parameters $\mu=1.142\ \text{kPa}$, $c_1=1.9074\ \text{kPa}$, $c_2=43.6$, and a quasi-planar fiber distribution with concentrations $a_1=0$, $b_1=12.36$, and orientations $\alpha_1=\beta_1=\gamma_1=0^{\circ}$. (a) Contour map of the maximum principal Cauchy stress (Pa) for the first six set of integration points; (b) plots of radial Cauchy stress $\sigma_{rr}$ versus coordinate $\vartheta$ along the circular path with radius $R=7\ \text{mm}$ for the models shown in (a). Since all $\sigma_{rr}(\vartheta)$ have a period of $\pi/2$, only the interval $0\leq\vartheta\leq\pi/2$ is shown. (c) Convergence rate of the DFD model to the isotropic solution. The error is evaluated in terms of average relative deviation of $\sigma_{rr}\left(\vartheta\right)$ from its mean value.}
	\label{fig:dfd_rotational_inv_test}
\end{figure}

As it can be observed in \cref{fig:dfd_rotational_inv_test}(a), showing the contour maps of the maximum principal Cauchy stress, the result obtained with low integration orders is strongly non-symmetric, exhibiting several directions of anisotropy. However, as the order increases, further directions of anisotropy develop, approaching to the limit a condition of perfect axial symmetry. To study the convergence rate to the isotropic solution, we analyzed the radial component $\sigma_{rr}$ of the Cauchy stress over the circular path of radius $R=7\ \text{mm}$. Due to the octahedral symmetry of the integration points, the profiles of the stress are $\pi/2$-periodic, and hence they are reported in \cref{fig:dfd_rotational_inv_test}(b) for the first six quadratures on the interval $0\leq\vartheta\leq\pi/2$. Because no analytical solution to this problem is available, we computed the error as the average relative deviation of $\sigma_{rr}\left(\vartheta\right)$ from its mean value, such that for the axisymmetric (isotropic) solution the error is exactly zero. The plot of convergence rate, reported in \cref{fig:dfd_rotational_inv_test}(c), shows a significant reduction of the relative error with $n$, and attains values lower than $1\%$ for $n\ge217$. The effective number of integration points, also reported in \cref{fig:dfd_rotational_inv_test}(c), was substantially reduced for all the quadratures.

\section{Applicative example: Z-plasty simulation}\label{sec:Z-plasty_analysis}
The traditional Z-plasty in skin surgery consists of three incisions of equal length. The central incision is aligned along the direction of the skin contraction, such as a linear scar or an extended burn scar. The other two incisions extend from either end of the central line in opposite directions at equal angles of $60^{\circ}$ to the central line, forming two identical, but opposed, triangular flaps. When these flaps are delaminated from the subcutaneous tissues, they are raised and transposed before suturing to create a new configuration resembling an inverted Z, where the central line becomes perpendicular to the original incision. This procedure increases the distance between the two ends of the initially vertical incision, thus releasing the contraction. The angles of the lateral incision can be adjusted to obtain the desired level of vertical elongation~\cite{Furnas1971}. Generally, larger angles result in greater elongation, but also higher localized deformations, usually correlated to the onset of complications like the distal necrosis~\cite{Gibson1967,Larrabee1984}.

In this section we analyze a set of Z-plasties with incision angle, or triangular flap amplitude, ranging from $20^\circ$ to $90^\circ$, representing an extension of the range in which these surgeries are usually performed in clinical procedures. To understand the influence of the collagen fibers, each set is analyzed for different mean fiber orientation $\alpha_1$. The results are then shown and discussed in terms of key mechanical quantities.

\subsection{FE model}
\begin{figure}[ht]
	\centering
	\includegraphics[width=18cm]{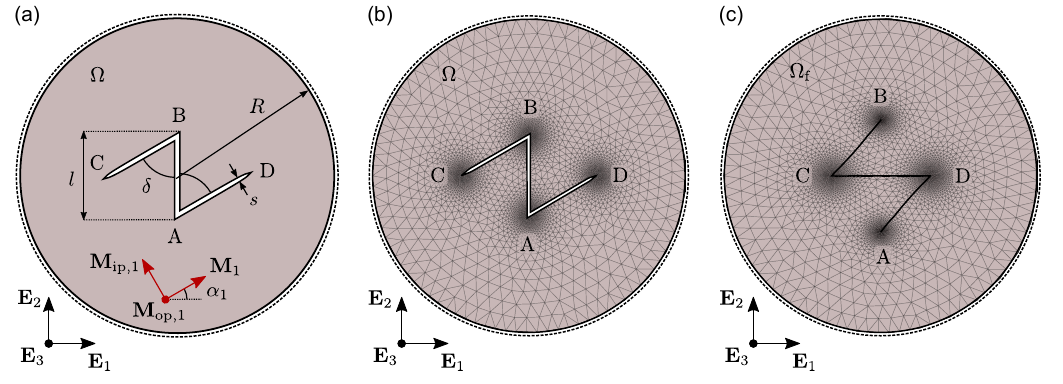}
	\caption{Representative scheme of the Z-plasty model. (a) Geometrical parameters; (b) discretized FE model for $\delta=60^\circ$ in the reference configuration, and (c) in the final configuration.}
	\label{fig:scheme_1ZP}
\end{figure}

With reference to the scheme reported in \cref{fig:scheme_1ZP}(a), we considered a Z-plasty with vertical and lateral incisions of length $l=50\ \text{mm}$ in a circular domain of radius $R=300\ \text{mm}$. The size for domain is not representative of real operations but is chosen to minimize boundary effects. For numerical purposes, the edges of the cuts were separated by a gap of $s=0.5\ \text{mm}$. The angles $\delta$ taken into consideration for the analysis were selected from the interval $20^\circ \le \delta \le 90^\circ$ at steps of $5^\circ$, leading to a set of $15$ models. Each domain in the study is discretized using three-node plane stress isoparametric elements (Abaqus CPS3 element), with a maximum size $h_{\rm max}=l/5$ in regions far from the incisions, progressively reducing to a minimum size $h_{\rm min}=l/150$ near areas of geometrical discontinuities, such as the cut tips. The outer boundary is assumed to be constrained in all directions. To simulate the closure and the suturing of the skin flaps, the nodes belonging to one side of the internal boundary are linked pairwise to the nodes on the other side, in a way similar to real stitches in surgical procedures. Using a kinematic multi-point constraint, the paired nodes are progressively joined, until the perfect coincidence is achieved. This approach spontaneously finds the final configuration of the suture, which is in general unknown before the analysis. Due to the complexity of adapting the mesh to the application of the multi-point constraints on the internal boundary, the FE models were generated using the algorithm described in~\citet{Alberini2021}. An example of generated domain for $\delta=60^\circ$ is reported \cref{fig:scheme_1ZP}(b).

Simulations were performed using the average parameters for human skin from \citet{Alberini2024a}, also used in \cref{sec:convergence_assessment}. In particular, thickness $t=2.92\ \text{mm}$, mechanical parameters $\mu=1.142\ \text{kPa}$, $c_1=1.907\ \text{kPa}$, $c_2=43.6$, and concentrations parameters $a_1=2.06$, $b_1=12.36$, were used. The fiber distribution was assumed to lie in the mean membrane plane, with out-of-plane angle $\beta_1=0^{\circ}$ and rolling angle $\gamma_1=0^{\circ}$. Concerning the mean in-plane fiber orientation, six different angles are considered, namely $\alpha_1=0^{\circ}$, $30^{\circ}$, $60^{\circ}$, $90^{\circ}$, $120^{\circ}$, $150^{\circ}$.

Considering the results of \cref{sec:convergence_assessment}, we used the Lebedev quadrature with $n=217$ for all the simulations, representing the minimum number of points with relative error below $1\%$ in the annular disk analysis.

\subsection{Results}
The fundamental measure to evaluate the performance of the Z-plasty is the vertical elongation strain $\varepsilon_{\rm v}$ parallel to the $\E{2}$ direction, defined as
\begin{equation}
	\varepsilon_{\rm v} = \dfrac{\parr{\Delta\x_{\rm v}-\Delta\X_{\rm v}}\cdot\E{2}}{\norm{\Delta\X_{\rm v}}}\eend,
	\label{eq:ZP_length_vertical_srain}
\end{equation}
where $\Delta\X_{\rm v}=\X_{\rm B}-\X_{\rm A}$ and $\Delta\x_{\rm v}=\x_{\rm B}-\x_{\rm A}$ are the distance vectors between the two points $\rm A$ and $\rm B$ of the scar in the reference and current configurations (see \cref{fig:scheme_1ZP}(a) and (b) for reference). The horizontal shortening strain $\varepsilon_{\rm h}$ is also considered, namely
\begin{equation}
	\varepsilon_{\rm h} = \dfrac{\parr{\Delta\X_{\rm h}-\Delta\x_{\rm h}}\cdot\E{1}}{\norm{\Delta\X_{\rm v}}}\eend,
	\label{eq:ZP_length_horiontal_srain}
\end{equation}
where $\Delta\X_{\rm h}=\X_{\rm D}-\X_{\rm C}$ and $\Delta\x_{\rm h}=\x_{\rm D}-\x_{\rm C}$ are the distance vectors between the two side points $\rm C$ and $\rm D$ of the surgery in the reference and current configurations (see \cref{fig:scheme_1ZP}(a) and (b) for reference). Note that $\varepsilon_{\rm h}$ is normalized with respect to $\norm{\Delta\X_{\rm v}}$, and not $\norm{\Delta\X_{\rm h}}$ as one would intuitively assume. The motivation behind this choice is that the initial positions $\X_{\rm C}$ and $\X_{\rm D}$ can vary between different surgeries as a function of the opening angle $\delta$ and the flap lengths, and normalizing with respect to $\norm{\Delta\X_{\rm h}}=\norm{\X_{\rm D}-\X_{\rm C}}$ would reduce the objectivity of $\varepsilon_{\rm h}$. Instead, $\norm{\Delta\X_{\rm v}}$ represents the initial scar length, which is constant for all the surgeries analyzed, hence enabling an appropriate comparison between the models.

\begin{figure}[t]
	\centering
	\includegraphics[width=18cm]{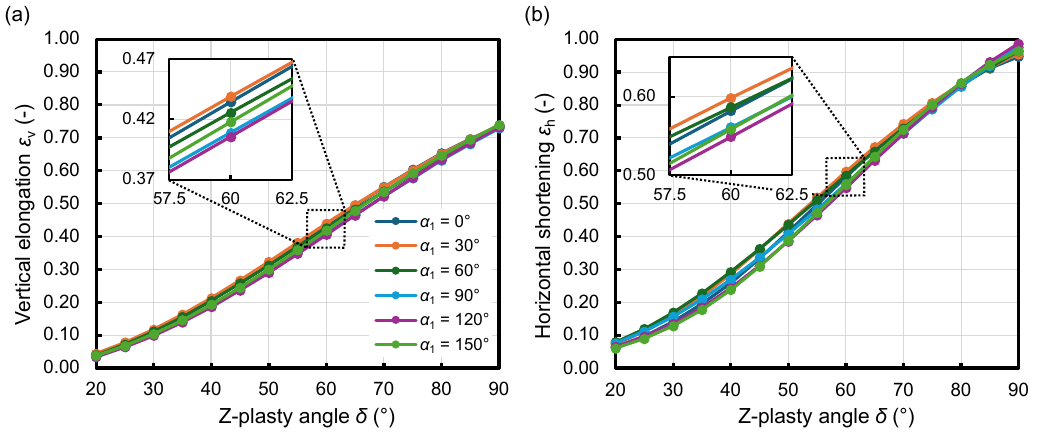}
	\caption{Vertical elongation and horizontal shortening strains in relation to the Z-plasty angle $\delta$ for the mean fiber direction angels $\alpha_1=0^{\circ}$, $30^{\circ}$, $60^{\circ}$, $90^{\circ}$, $120^{\circ}$, $150^{\circ}$. (a) Vertical elongation strain $\varepsilon_{\rm v}$; (b) horizontal shortening strain $\varepsilon_{\rm h}$. Magnifications are shown for the special case of $\delta=60^{\circ}$ Z-plasty.}
	\label{fig:ZP_V_and_H_strains}
\end{figure}

The two strain measures are reported in \cref{fig:ZP_V_and_H_strains} for all the fiber orientations $\alpha_1$ analyzed. The vertical elongation strain $\varepsilon_{\rm v}$ in \cref{fig:ZP_V_and_H_strains}(a) is always greater than zero and increases almost linearly with $\delta$, ranging from $3.3\div4.4\%$ at $\delta=20^{\circ}$ to $79.2\div73.9\%$ at $\delta=90^{\circ}$. This indicates that the surgical procedures consistently produce a lengthening effect.
Concerning the horizontal shortening strain $\varepsilon_{\rm h}$, shown in \cref{fig:ZP_V_and_H_strains}(b), the trends are analogous to the vertical elongation in \cref{fig:ZP_V_and_H_strains}(b), showing an approximately linear trend for all the angles $\alpha_1$. The values span from $6.1\div8.0\%$ at $\delta=20^{\circ}$ to $94.8\div98.7\%$ at $\delta=90^{\circ}$.
Interestingly, both the strains seem to be almost unaffected by the mean fiber direction $\alpha_1$, with $\varepsilon_{\rm v}$ and $\varepsilon_{\rm h}$ showing only $2.4\%$ and $3.8\%$ difference between maximum and minimum values on average, respectively. However, it is worth noticing that the vertical elongation $\varepsilon_{\rm v}$ is always lower than the the horizontal shortening $\varepsilon_{\rm h}$ as the Z-plasty angle $\delta$ varies. This trend becomes particularly evident at higher angles, where the initial distance between points $\rm C$ and $\rm D$ in the reference configuration significantly exceeds the scar length $l$ (see geometrical scheme in \cref{fig:scheme_1ZP}(a)). Following the transposition, points $\rm C$ and $\rm D$ align along the central limb, corresponding to the joined sides of the two triangular flaps initially sized $l$, meaning that $\norm{\Delta\x_{\rm h}}\approx l$. Consequently, the deformation $\varepsilon_{\rm h}$, being proportional to $\parr{\Delta\X_{\rm h}-\Delta\x_{\rm h}}\cdot\E{1}\approx\norm{\Delta\X_{\rm h}}-\norm{\Delta\x_{\rm h}}\approx\norm{\Delta\X_{\rm h}}-l$, attains high values when the initial distance $\norm{\Delta\X_{\rm h}}$ between points $\rm C$ and $\rm D$ is much larger than $l$. Clinically, while higher strains $\varepsilon_{\rm v}$ are desirable, the horizontal strains $\varepsilon_{\rm h}$ should be reduced as much as possible in order to reduce stresses. Therefore, selecting higher angles $\delta$ to increase the elongation is a choice that must be taken carefully.

\begin{figure}[p]
	\centering
	\includegraphics[width=18cm]{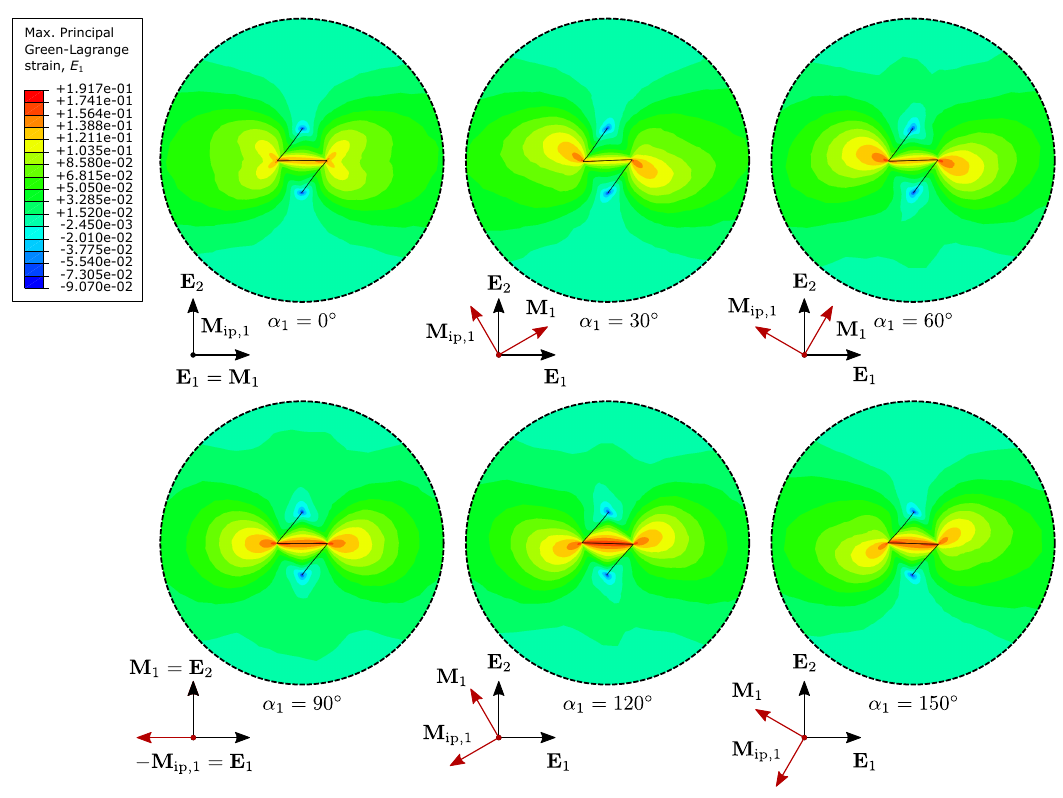}
	\caption{Maps of the maximum principal in-plane component $E_{1}$ $(>E_{2},E_{3})$ of the Green-Lagrange strain tensor in the final configuration, for the $\delta=60^{\circ}$ Z-plasty with the different mean in-plane fiber orientations $\alpha_1$. Maps are cropped to $R=150\ \text{mm}$ to improve readability.}
	\label{fig:ZP_princ_strain_map}
\end{figure}

The influence of anisotropy can be clearly appreciated from the distribution of the strains in the final configuration. \Cref{fig:ZP_princ_strain_map} illustrates the map of the maximum principal in-plane component $E_{1}$ of the Green-Lagrange strain tensor $\mathbf{E}=1/2\parr{\C-\I}$ for the different mean fiber directions $\alpha_1$ considered. Due to the anisotropy of the material, the lobes lateral to the points $\rm C$ and $\rm D$ tend to align with the direction $\Mipi$ perpendicular to the fibers (red to yellow shades in \cref{fig:ZP_princ_strain_map}). This happens because the material tends to redistribute the increased horizontal strains, concentrated across the middle limb, toward the neighboring regions in the direction with lower stiffness.

\begin{figure}[t]
	\centering
	\includegraphics[width=14cm]{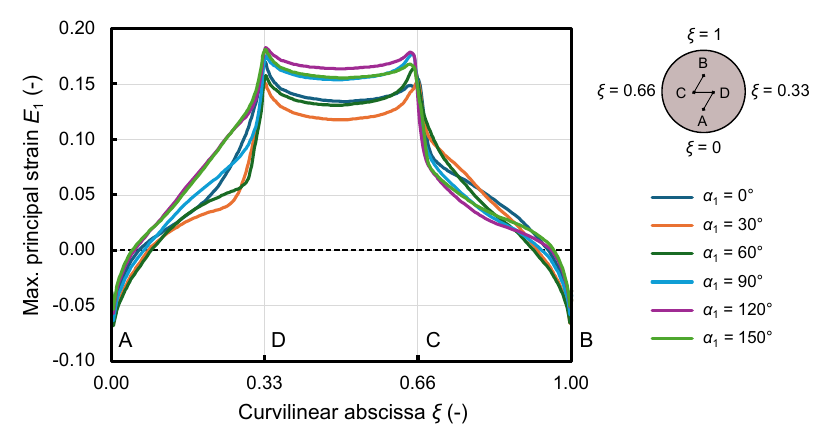}
	\caption{Maximum principal in-plane strain component $E_{1}$ along the suture for the Z-plasty case $\delta=60^{\circ}$. The curvilinear abscissa $\xi=L/L_{s}$, $\xi\in\pars{0,1}$ represents the distance $L$ from the point $\rm A$ at the lower end of the suture normalized with respect to the global length of the suture $L_{s}$ in the final configuration.}
	\label{fig:ZP_ZP60_Eip_suture}
\end{figure}

A representative profile of the maximum principal in-plane strain $E_{1}$ along the suture is depicted in \cref{fig:ZP_ZP60_Eip_suture} for $\delta=60^{\circ}$. The curves exhibit negative strains in the proximity of the suture ends, corresponding to the points $A$ and $B$, while two peaks emerges close to the geometrical singularities located at points $\rm C$ and $\rm D$. Between the latter, the strain remains high, indicating that the central limb is significantly more stressed than the others.

\begin{figure}[t]
	\centering
	\includegraphics[width=18cm]{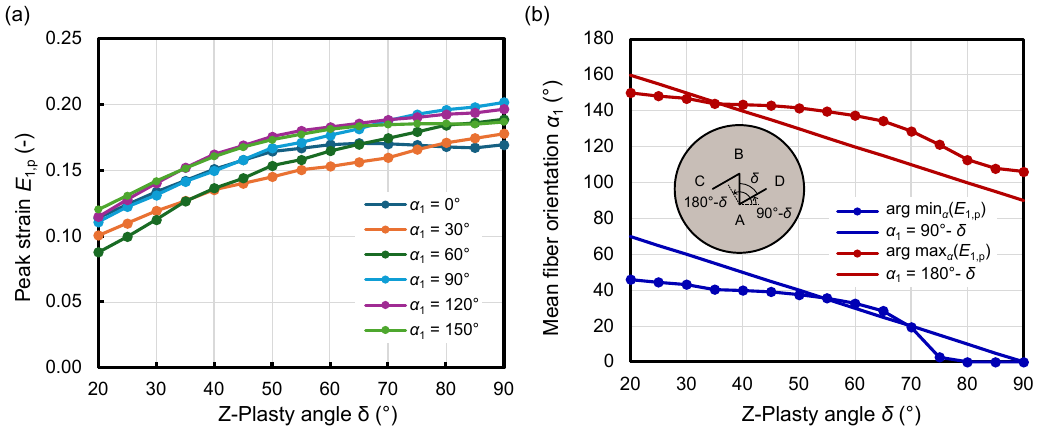}
	\caption{Peak strain analysis. (a) Peak strain $E_{1,\mathrm{p}}$ along the suture for the different Z-plasty angle $\delta$ and mean fiber direction $\alpha_1$. (b) Mechanical and geometrical relationships between the angles $\delta$ and $\alpha_1$. The angles $\alpha_{1,\mathrm{min}}$ and $\alpha_{1,\mathrm{max}}$, minimizing and maximizing $E_{1,\mathrm{p}}$ respectively, are obtained from the third degree polynomial fitting of the plots in (a) along $\alpha_1$ for every angle $\delta$.}
	\label{fig:ZP_peak_strain}
\end{figure}

\Cref{fig:ZP_peak_strain}(a) shows the peak principal strains $E_{1,\mathrm{p}}$ from all the maximum strain $E_{1}$ profiles in relation to the Z-plasty angle $\delta$. The plots, starting from strains of $8.8\div12.0\%$ at $\delta=20^{\circ}$, tend to increase with decreasing slope, ending at values of $17.0\div20.2\%$ at $\delta=90^{\circ}$. Even though the curves seem to overlap randomly, they reveal a correlation between the mean fiber orientation $\alpha_1$ and the flap angle $\delta$. By observing the plots in \cref{fig:ZP_peak_strain}(b) showing the angles $\alpha_1$ that minimize $E_{1,\mathrm{p}}$, we observe that the minimum peak strain occurs when the fibers in the reference configuration are approximately parallel to the lateral incisions $\rm{AD}$ and $\rm{BC}$, i.e. when $\alpha_1=90^{\circ}-\delta$. In fact, the sides of the triangular flaps initially laying on the lateral incisions occupy the central limb once the suture is closed. Therefore, when the conditions $\alpha_1=90^{\circ}-\delta$ is achieved, the fibers can exert the maximum stiffness, thus reducing the strain. Conversely, when the fibers lay at right angle to the lateral limbs in the reference configuration the triangular flap has lower stiffness along the central limb, leading to increased strains.

\section{Discussion}\label{sec:discussion}
Discrete fiber dispersion models are approximation of the continuous fiber dispersion approach, offering comparable versatility for modeling the complex mechanical behavior of general fiber distributions at lower computational cost. The discrete approach enables a straightforward exclusion of fibers under compression by means of the single-fiber strain-energy in \cref{eq:Psi_single_fiber}. As a result, the mechanical response is governed solely by the integration points $\N_{j}$ under tension and the ground matrix subjected to arbitrary finite deformations.

In this study, considering the relevant works of \citet{Ehret2010, Skacel2015} which compared different integration schemes, we studied a DFD model with Lebedev quadrature. The model has been implemented in the commercial FE software Abaqus for general three-dimensional formulation as well as for plane stress problems. The main advantage of the Lebedev quadrature is that the integrations points $\N_{j}$ are built on the unit sphere leveraging the octahedral symmetries, so that a uniform distribution of spherical point grid is obtained. In addition, the non-negativity of weights $w_j$ preserves the poly-convexity of the strain-energy function in \cref{eq:Psi_DFD}.

To reduce the computational effort, we limited the integration to one hemisphere by exploiting the fiber symmetry $\rho_{i}\parr{\N}=\rho_{i}\parr{-\N}$. Additionally, we excluded directions $\N_{j}$ with negligible influence, specifically those with values $\varrho_{i,j}$ smaller than $10^{-6}$. This operation sensibly reduces the number of integration points, especially when concentrations parameters $a_i$ and $b_i$ increase.
However, the sharpness of the distribution $\rho_{i}$ is limited by the quadrature order. In fact, quadrature schemes are designed to exactly integrate all the spherical harmonics up to a determined and finite degree. Therefore, when the integrand function cannot be approximated by a combination of harmonics with degree lower than the quadrature order the accuracy of the result is not guaranteed. In the limit case where the PDF becomes singular, discrete integrations can no longer be used. Recalling the integrand in \cref{eq:Psi_DFD}, and the in-plane and out-of-plane distributions in \cref{eq:fib_vonMises_ip,eq:fib_vonMises_op}, respectively, this happens when either $a_i$ and $b_i$ approach zero. In particular, $\rho_i(\N)$ degenerates to a Dirac delta function in the direction of $\Mi$ for $a_i,b_i\rightarrow\infty$, $\rho_i(\N)=\delta\parr{\left\|\N-\Mi\right\|}$, or to a singular planar distribution in the plane $\{\Mi,\, \Mipi\}$ for $b_i\rightarrow\infty$, $\rho_i(\N)=\rho_{\ip,i}\parr{\N}\delta\parr{\N\cdot\Mopi}$.

One way to address this issue is by setting a limit for the parameters $a_i$ and $b_i$, beyond which the fiber distribution can be considered perfectly aligned, and the DFD model can be replaced by the simpler \citet{Holzapfel2000} model. Likewise, when only $b_i$ exceeds the limit, the distribution can be assumed as perfectly planar, $\rho_i(\N)=\rho_{\ip,i}$, and \cref{eq:Psi_DFD} can be reduced to an integral over the circle in the plane $\{\Mi,\, \Mipi\}$ and solved using a simple univariate Gauss quadrature. \citet{Skacel2015} proposed to limit the concentration to $b_i\leq20$ using $n=160$ integration points, but they considered a transversely isotropic fiber distribution only. For a comprehensive analysis, more values of $a_i$ and $b_i$ should be explored, taking also into account more complex loading configurations. However, this type of investigation goes beyond the scope of our work, which was focused on assessing the computational performance of the Lebedev quadrature in applications related to large deformation problems.

Another cause of discontinuity of the integrand function is the compression switch incorporated in the single-fiber strain-energy shown in \cref{eq:Psi_single_fiber}. By juxtaposing the convergence rates obtained in \cref{fig:dfd_uniaxial_test}(b) and \cref{fig:dfd_shear_test}(b), it seems that the simple shear test, despite most of the fibers were loaded in compression, converges with a rate similar to the uniaxial test, in which most of the fibers are loaded in tension. Therefore, we can conclude that the discontinuity related to the compression switch does not significantly impact the integration accuracy.

The analysis of the annular disk torsion in \cref{sec:annular disk}, as predicted by the observations of \citet{Ehret2010}, revealed that the discrete integration can lead to strongly anisotropic results for isotropic fiber distributions if an insufficient number of integration points is used. Such anisotropy arises from the intrinsic limitation of quadrature rules in exactly integrating a function under general rigid rotations of the integration points. The Lebedev quadrature rule, unlike other schemes, is invariant under the octahedral rotation group, but not for general rotations. Nonetheless, the isotropic disk approaches the expected axially symmetric solution for increasing number of integration points. The rapid convergence shown in \cref{fig:dfd_rotational_inv_test}(c) is likely due to the uniform distribution of the points as depicted in \cref{fig:Lebedev_points}. This suggests that similar quadratures based on higher polyhedral symmetries, such as that proposed by \citet{Ahrens2009}, could converge even faster than Lebedev's.

Based on the convergence analyses presented in \cref{sec:convergence_assessment} for concentrations $0\leq a_1\leq2.06$ and $b_1=12.36$, it turned out that the Lebedev quadrature provides accurate results using the set with $n=217$ points on the hemisphere. This number can be further reduced by neglecting the points with $\varrho_{i,j}=w_{j}\rho_{i}(\N_{j})<10^{-6}$, without significantly altering the accuracy.

Once defined the required number of integration points, we included the DFD model with Lebedev quadrature in a simulation of a corrective skin surgical technique, and analyzed its mechanical properties. In particular, we simulated a Z-plasty for different geometrical configurations of the initial incision, varying the angle $\delta$ between the vertical and lateral cuts from $20^\circ$ to $90^\circ$. Moreover, we explored the influence of the material anisotropy considering different orientation of the mean fiber direction $\Mi$ in the plane $\{\E{1}, \E{2}\}$ by changing the angle $\alpha_1$.

The analyses revealed that the vertical elongation $\varepsilon_{\rm v}$ and the horizontal shortening $\varepsilon_{\rm h}$, see \cref{fig:ZP_V_and_H_strains}, are almost unaffected by the mean fiber orientation. Moreover, the relation with the Z-plasty angle $\delta$ is almost linear for both the measures. The strains spans in the interval $20^{\circ}\leq\delta\leq90^{\circ}$ from $3.3\div4.4\%$ to $79.2\div73.9\%$ for $\varepsilon_{\rm v}$, and from $6.1\div8.0\%$ to $94.8\div98.7\%$ for $\varepsilon_{\rm v}$. Interestingly, these results are in agreement with our preliminary isotropic analysis in \citet{Spagnoli2022a}, in which the strain $\varepsilon_{\rm v}$ ranged from $\sim5\%$ to $\sim73\%$ for the same investigated domain of the angle $\delta$. This suggest that the Z-plasty has some intrinsic geometrical properties independent of the material anisotropy, at least in the range of parameters considered. This result appears particularly useful for the surgical practice, since it simplifies the selection of the angle $\delta$ depending on the desired $\varepsilon_{\rm v}$ to achieve, and regardless the material properties.

The effects of anisotropy, however, can be observed in the map of the maximum principal Green-Lagrange strain $E_{1}$ in \cref{fig:ZP_princ_strain_map} and in its profile along the suture in \cref{fig:ZP_ZP60_Eip_suture} for the representative Z-plasty with $\delta=60^{\circ}$. The strain $E_{1}$ is negative near the suture ends and reaches its peak at points $\rm C$ and $\rm D$. The central limb experiences higher strain, indicating a higher stress level compared to that in the lateral limbs. The analysis of the peak principal strain $E_{1,\mathrm{p}}$ in \cref{fig:ZP_peak_strain}(a) revealed a relationship between the mean fiber orientation $\alpha_1$ and the Z-plasty angle $\delta$. In particular, the minimum strain $E_{1,\mathrm{p}}$ occurs when the fibers run parallel to the lateral incisions $\rm{AD}$ and $\rm{BC}$ in the initial configuration, i.e. $\alpha_1=90^{\circ}-\delta$, while the maximum occurs when the fibers are perpendicular to the lateral incisions, i.e. $\alpha_1=180^{\circ}-\delta$.

The results herein presented, however, are limited to only one set of mechanical and microstructural parameters averaged from test on human abdominal skin~\cite{Alberini2024a}. To demonstrate whether the Z-plasty has intrinsic geometrical properties, different set of parameters should be considered, analyzing the influence of both the mechanical and microstructural parameters. This type of analysis, such as that performed by \citet{Stowers2021} on various skin surgical procedures, could also reveal further relationships between the geometry and the fiber orientation, enabling a deep understanding of this surgical procedure. This could also pave the way for the development of innovative patient-oriented surgical strategies.

\section{Conclusions}\label{sec:conclusions}
In this work we analyzed the computational efficiency of a Discrete Fiber Dispersion (DFD) model integrated with the Lebedev quadrature rule. The model has been implemented in the commercial FE software Abaqus for perfectly incompressible materials in plane stress formulation. The model effectively excludes fibers under compression and provides accurate results using $n=217$ integration points.

The DFD model was then used for a parametric analysis of the mechanical behavior of the Z-plasty surgery. The results showed that the mean fiber direction has little effects on the Z-plasty lengthening outcome, but significantly influences the maximum strain measured in proximity of the suture. Specifically, the maximum strain is minimized when the mean fiber direction is parallel to the lateral cuts in the reference configuration. These results, although limited to one set of mechanical and microstructural parameters of human skin, provided insight about the mechanical performances of the Z-plasty, strengthening the understanding of this widespread corrective surgery.



\bibliographystyle{elsarticle-num-names}
\biboptions{sort&compress}
\bibliography{zot_library}

\end{document}